\documentclass[fleqn,usenatbib]{mnras}

\usepackage{newtxtext,newtxmath}
\usepackage[T1]{fontenc}
\usepackage{csquotes}
\usepackage{xcolor}

\DeclareRobustCommand{\VAN}[3]{#2}
\let\VANthebibliography\thebibliography
\def\thebibliography{\DeclareRobustCommand{\VAN}[3]{##3}\VANthebibliography}

\usepackage{graphicx}	
\usepackage{amsmath}	
\usepackage{multirow}

\usepackage{array}
\newcolumntype{L}[1]{>{\raggedright\let\newline\\\arraybackslash\hspace{0pt}}m{#1}}
\newcolumntype{C}[1]{>{\centering\let\newline\\\arraybackslash\hspace{0pt}}m{#1}}

\newcommand{\astro}{\emph{AstroSat}}
\newcommand{\rxte}{\emph{RXTE}}
\newcommand{\chan}{\emph{Chandra}}
\newcommand{\suz}{\emph{Suzaku}}
\newcommand{\xmm}{\emph{XMM}-Newton}
\newcommand{\xte}{XTE J1710--281}

\title[Orbital Period Glitches in \xte]{Eclipse Timings of the LMXB \xte\ : Discovery of a third orbital period glitch}

\author[C. Jain, R. Sharma and B. Paul]{
Chetana Jain,$^{1}$\thanks{E-mail: chetanajain11@gmail.com (CJ)}
Rahul Sharma,$^{2}$\thanks{E-mail: rsharma@rri.res.in (RS)}
and Biswajit Paul$^{2}$
\\
$^{1}$Hansraj College, University of Delhi, Delhi 110007, India\\
$^{2}$Raman Research Institute, C.V. Raman Avenue, Bangalore 560080 Karnataka, India \\
}

\date{Accepted XXX. Received YYY; in original form ZZZ}

\pubyear{2022}

\begin{document}

\label{firstpage}
\pagerange{\pageref{firstpage}--\pageref{lastpage}}
\maketitle

\begin{abstract}

We present an updated measurement of orbital period evolution of LMXB \xte\ by using eclipse timing technique. Using data obtained with \xmm, \suz, \rxte, \chan\ and \astro\ observatories, we report 21 new measurements of X-ray mid-eclipse times. We have discovered a third orbital period glitch in \xte\ with an F-test false alarm probability of $\sim$0.7\% for occurrence of the third glitch and report detection of four distinct epochs of orbital period in this system. This work presents a more robust estimation of occurrence of the second orbital period glitch. However, the epoch of occurrence of the third glitch is poorly constrained, between MJD 55726 to 56402. We have put lower limits of 1.48 ms, 0.97 ms and 0.45 ms, on sudden changes in orbital period between the successive epochs. We discuss the implications of our findings in context of magnetic nature of the companion star and possible scattering events with circum-binary objects around this binary system.

\end{abstract}




\section{Introduction}

The orbital period is one of the most fundamental parameter that characterizes a binary system. Therefore, in order to understand and constrain the properties of the binary components, astronomers have extensively studied the evolution mechanism of the orbital period. Several measurements have been done for both Low Mass X-ray Binaries (LMXBs) and High Mass X-ray Binaries (HMXBs) with observations spanning up to a few decades. The measurement techniques include pulse timing technique (LMXBs: \citealt{Deeter91, Jain07, Staubert09}; HMXBs:  \citealt{Naik04, Mukherjee06, Baykal06, Raichur10, Jenke12}), eclipse timing technique (LMXBs: \citealt{Wolff09, Jain10, Jain11, Jain17, Ponti17}; HMXBs: \citealt{Falanga15, Islam16}), use of stable orbital intensity profile as a time marker \citep{Chou01, Singh02, Peuten14} and measurement of the Doppler shift in the spectrum of the companion star \citep{Gonzalez14}. 

The orbits of X-ray binaries are expected to evolve due to mass transfer and re-distribution of the angular momentum arising from the interaction of the binary components \citep{Heuvel94}, mass loss from the binary system due to processes such as, jet emission from the compact object, irradiative evaporation of the secondary star or in the form of accretion disc winds \citep{Ruderman89, Brookshaw93, Ponti12}, tidal interactions between the binary components \citep{Lecar76, Zahn77} and loss of orbital angular momentum via gravitational wave radiation or magnetic braking of the tidally coupled primary \citep{Rappaport83, Applegate92, Verbunt93}. 

As a result of these mechanisms, the orbital separation in X-ray binaries can increase \citep{Homer98, Parmar00, Jain07, Jain10} or decrease \citep{Deeter91, Klis93, Paul04}. However, it has been observed that in most of the LMXBs, the orbital period is increasing at a rate much higher than that predicted by a conservative mass transfer or by gravitational wave radiation \citep{Jain07, Jain10, Sanna16}. The orbital decay in X-ray binary systems is also unusual and is much faster than that predicted by conventional methods of gravitational radiation, magnetic braking and mass loss from the system \citep{Peuten14}. In addition, there are two LMXB systems that show sudden changes in the orbital period \citep{Wolff09, Jain11}. Another interesting LMXB is MXB 1658-298, which shows orbital period decay on a timescale spanning about four decades. But on shorter timescales, there are indications of presence of a third body around the binary system \citep{Jain17}. 

The object of this research work is LMXB \xte\ which was discovered with Rossi X-ray Timing Explorer (\rxte) in 1998 \citep{Markwardt98}. It is located at a distance of $\sim$15 kpc and has an inclination of about 80$^{\circ}$ \citep{Frank87, Markwardt01}. The compact object in this binary system is a neutron star \citep{Markwardt01}. The orbital period of \xte\ is 3.28 h. This source exhibits dipping activity and complete, sharp eclipse transitions \citep{Markwardt98, Markwardt01}. The eclipse phase lasts for a duration of about 420 s \citep{Jain11}. From observations spanning more than a decade, \citet{Jain11} have detected the presence of three distinct epochs of orbital period. However, due to non-availability of sufficient data points at that time, the time of occurrence of second orbital glitch was uncertain. Therefore, for this work, we have extended the time-base for \xte\ observations and present the updated results of eclipse timing analysis. 

\section{Observations}

Data for the present analysis were obtained from observations made with instruments on-board \xmm, \suz, \rxte, \chan\ and \astro. Table \ref{table:log} lists the log of observations used in the present work.

\begin{table*}
\centering
\caption{Log of observations and mid eclipse time measurements of \xte.}
\label{table:log}
\begin{tabular}{l l l l l l l}
\hline
Mission--           & Observation    & Date of Observation   & Exposure  &    Orbital     & Mid-eclipse Time  &   Uncertainty\\
Instrument          &  Id            &  (DD-MM-YYYY)         & (ks)      &    Cycle*             & MJD (d)           & 1$\sigma$ (d)\\
\hline
\xmm-PN     & 0206990401        & 22-02-2004    & 13.9      & 13214            & 53057.423293     &  0.000015\\
\suz-XIS    & 404068010         & 23-03-2010    & 76        & 29472            & 55280.070265    & 0.000005 \\
\rxte-PCA   & 94314-01-07-03    & 10-10-2010    & 9.2       & 30933 &       55479.805016&        0.000016\\
\rxte-PCA   & 94314-01-07-03    & 10-10-2010    & 9.2       & 30934 &       55479.941720 &       0.000011\\
\rxte-PCA   & 94314-01-08-00    & 31-10-2010    & 6.9       & 31084 &       55500.448340 &       0.000015 \\
\rxte-PCA   & 94314-01-09-01    & 08-11-2010    & 6.7       & 31142 &       55508.377619 &       0.000052\\
\rxte-PCA   & 94314-01-10-00    & 13-01-2011    & 15.4      & 31627 &       55574.682446 &       0.000011\\
\rxte-PCA   & 94314-01-10-00    & 13-01-2011    & 15.4      & 31628 &       55574.819127 &       0.000017\\
\rxte-PCA   & 94314-01-11-00    & 23-04-2011    & 13.6      & 32357 &       55674.481414 &       0.000017\\
\rxte-PCA   & 94314-01-11-01    & 24-04-2011    & 6.9       & 32365 &       55675.575093 &       0.000017\\
\rxte-PCA   & 96329-01-01-00    & 06-07-2011    & 14        & 32901 &       55748.852182 &       0.000017\\
\chan-ACIS  & 12468             & 23-07-2011    & 75        & 33026	&      55765.941025  &      0.000018\\
\rxte-PCA   & 96329-01-02-01    & 24-07-2011    & 1.8       & 33028 &       55766.214455 &       0.000017\\
\rxte-PCA   & 96329-01-02-00   & 24-07-2011    & 19.7       & 33030 &       55766.487866 &      0.000011\\
\rxte-PCA   & 96329-01-02-00    & 24-07-2011    & 19.7      & 33031 &       55766.624630 &       0.000011\\
\rxte-PCA   & 96329-01-03-000   & 05-08-2011    & 25        & 33115 &       55778.108319 &      0.000011\\
\chan-ACIS  & 12469             & 07-08-2011    & 75        & 33135	&      55780.842543	 &     0.000013\\
\astro-LAXPC    & 9000001188    & 19-04-2017    & 11        & 48361     &   57862.404031     &   0.000040 \\
\astro-LAXPC    & 9000001188    & 19-04-2017    & 11        & 48362     &   57862.540720     &   0.000046 \\
\astro-LAXPC    & 9000001382    & 14-07-2017    & 12.5      & 48991     &   57948.531930    &   0.000017\\
\astro-LAXPC    & 9000001382    & 14-07-2017    & 12.5      & 48992     &   57948.668632    &   0.000017\\
\hline
\multicolumn{7}{l}{$*$ w.r.t. MJD 51251.061141 \citep{Jain11}.} \\
\end{tabular}
\end{table*}

\xmm\ \citep{Jansen01} consists of three focal plane instruments (EPIC: European Photon Imaging Camera) - one pn-CCD camera \citep{Struder01} and two MOS detectors \citep{Turner01}. \xte\ was observed with \xmm\ in 2004 for an exposure of about 14 ks (Table \ref{table:log}).
The observation data files were processed using the XMM Science Analysis System (SAS version 20.0.0). For this work, 0.5--10 keV EPIC-pn imaging mode data has been used. The X-ray events were extracted from a circular region of radius 45 arcsec centered on the source position. The background events were extracted from a source-free circular region of same radius as the source. The background subtracted light curve was barycenter corrected using the SAS tool \texttt{barycen}.

The X-ray Imaging Spectrometer (XIS) on-board \suz\ \citep{Mitsuda07} consists of four units and covers the energy range of 0.2--12 keV \citep{Koyama07}. Of the four XIS units, three are front-illuminated CCDs (XIS0, XIS2, and XIS3) and one is back-illuminated CCD (XIS1). 
For the current analysis, we have utilized the data from XIS0 and XIS3 detectors. \xte\ was observed by \suz\ in 2010 for a duration of about 190 ks. The net exposure time was about 76 ks (Table \ref{table:log}). XIS detectors were operated in the standard data mode with normal window operation that provided a timing resolution of 8 s. The unfiltered event files were processed with the CALDB version 20181010 by using \texttt{aepipeline}. The XIS event files did not suffer from photon pile-up \citep{Raman18, Sharma20b}. We combined the cleaned event files for both $3 \times 3$ and $5 \times 5$ pixel mode for each CCD, by using the \texttt{xselect}. The combined cleaned events were corrected for the solar system barycenter using \texttt{aebarycen}. The clean event files were then used to extract the 0.5--10 keV light curve from a circular region of 3.5 arcmin centered on the source. An annular region of inner (outer) radius of 4.5 arcmin (6.5 arcmin) around the source was selected for the background. The light curves from XIS0 and XIS3 were added using \texttt{lcmath}. 

The \rxte-Proportional Counter Array (PCA) consists of an array of five collimated proportional counter units having a total photon collection area of 6500 cm$^2$ \citep{Jahoda96,Jahoda06}. For the current work, we have used all the \rxte\ observations subsequent to our previous compilation of mid-eclipse times \citep[Table \ref{table:log},][]{Jain11}. The \rxte-PCA data was collected in the Good Xenon mode. The 2--20 keV light curves were generated by using the \texttt{ftool-seextrct}. The background counts were estimated by using the \texttt{ftool-pcabackest}, assuming a faint source model. The background subtracted light curves were corrected for solar system barycenter by using the \texttt{ftool-faxbary}.

\xte\ was observed with the \chan\- Advanced CCD Imaging Spectrometer (ACIS) detector \citep{Weisskopf02, Garmire03} in 2011 for an exposure of $\sim$75 ks (Table \ref{table:log}). The data was collected from ACIS-S CCDs in the timed exposure mode having a time resolution of $\sim 1.74$ s. The Chandra Interactive Analysis of Observations (CIAO) software version 4.14 was used to generate level 2 files by using \texttt{chandra\_repro}. 
The 0.5--8 keV light curves were extracted from a circular region of radius 10 arcsec. A region of similar radius away from source was used to obtain the background events. Barycenter correction was done using \textsc{CIAO} tool \texttt{axbary}.

\astro\ is India's first multi-wavelength astronomy mission \citep{Agrawal06, Singh14} and the Large Area X-ray Proportional Counter (LAXPC) is one of its primary payloads \citep{Yadav16, Agrawal17}. It consists of three co-aligned proportional counters (LAXPC10, LAXPC20 and LAXPC30) which are sensitive to the X-ray photons in the 3--80 keV energy range, with a total effective area of 6000 cm$^{-2}$ at 15 keV. We have used Event Analysis (EA) mode data from LAXPC10 and LAXPC20 for the present work. Data from LAXPC30 was not used due to high background and gain variations of the instrument \citep{Agrawal17, Antia17}. In order to minimize the contribution of background in our analysis, we used data from the top layers (L1, L2) only \citep[also see,][for details]{Beri2019, Sharma20}.
The level 1 data were processed by using the standard LAXPC software  (\textsc{LaxpcSoft}: version 3.4.2)\footnote{\url{https://www.tifr.res.in/~astrosat\_laxpc/LaxpcSoft.html}}. The 3--15 keV source and background light curves were extracted by using the tool \texttt{laxpcl1}. The photon arrival times in level 2 files were corrected to the solar system barycenter by using \texttt{as1bary}\footnote{\url{http://astrosat-ssc.iucaa.in/?q=data\_and\_analysis}} tool. The barycenter and background corrected light curves from LAXPC10 and LAXPC20 were added using \texttt{lcmath}.

Barycenter correction for all the light curves has been done using the JPL DE--405 ephemeris. The source coordinates used for this conversion were R.A. = 17$^h$ 10$^m$ 12.3$^s$ and and dec = -28$^\circ$ 07$^\prime$ 54$^{\prime\prime}$ \citep{Ebisawa03}.

\section{ANALYSIS AND RESULTS}

\begin{figure*}
\centering
\begin{minipage}{.32\linewidth}
\centering
\includegraphics[width=1.2\linewidth]{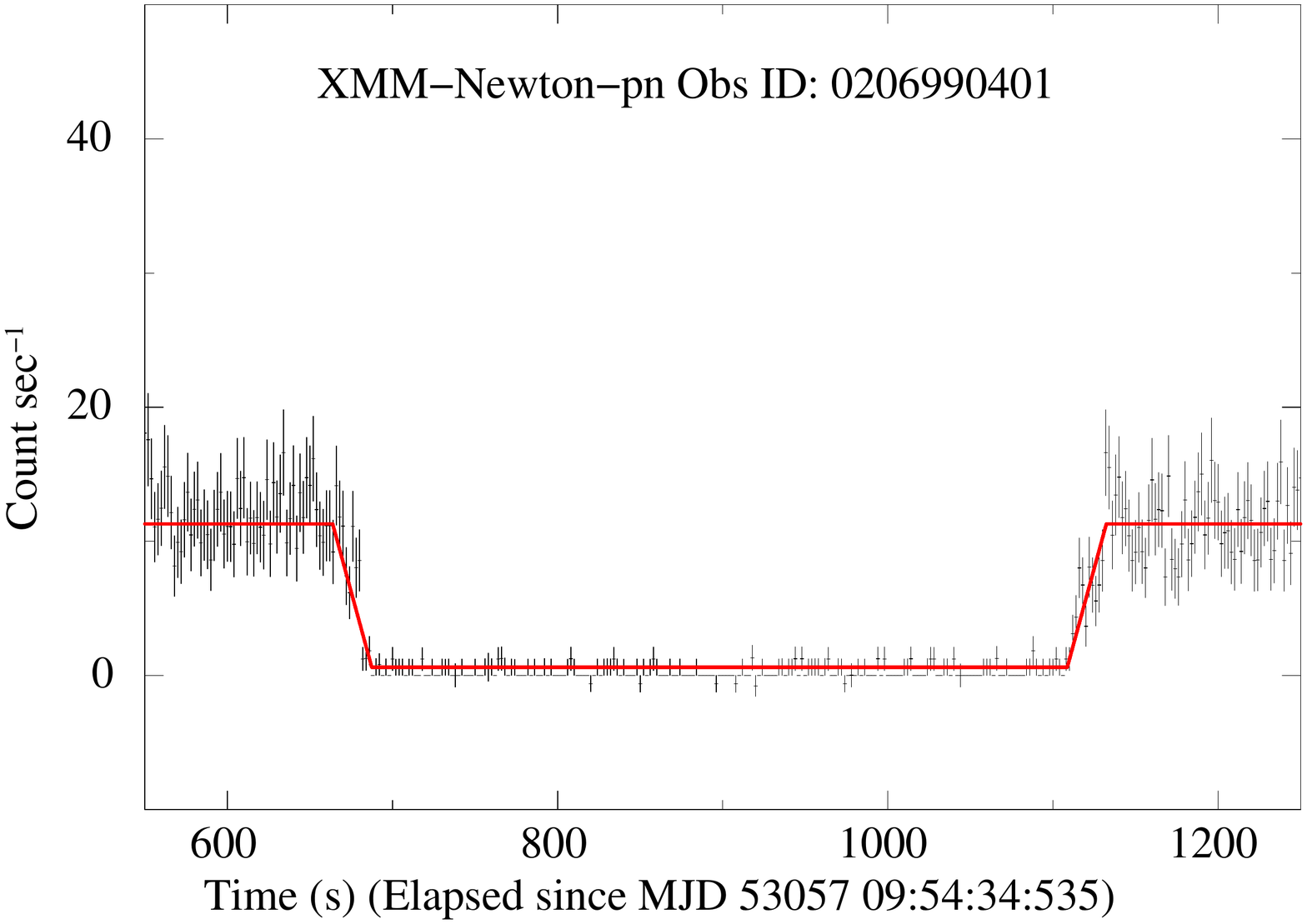}
\end{minipage}
\begin{minipage}{.32\linewidth}
\centering
\includegraphics[width=1.2\linewidth]{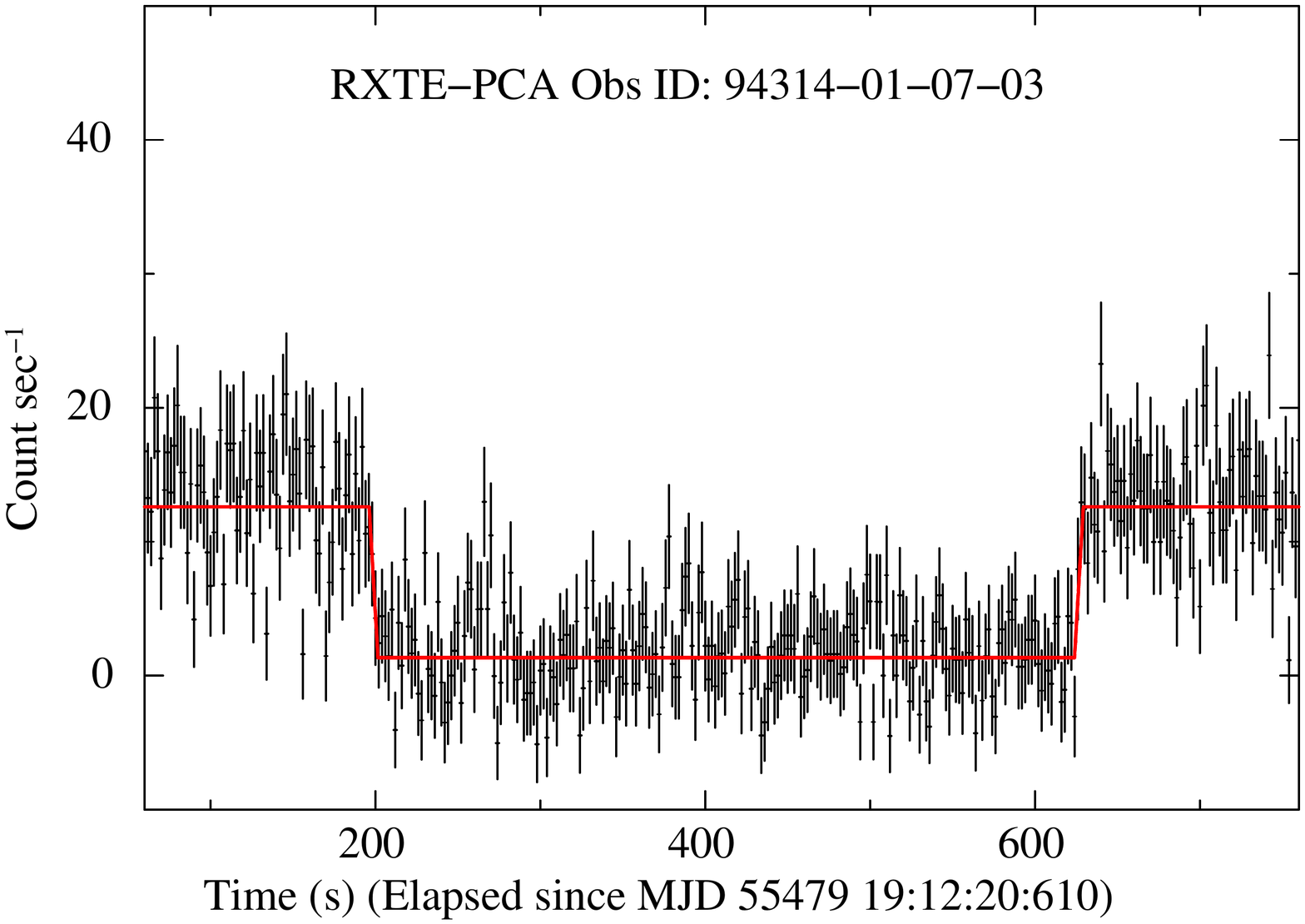}
\end{minipage}
\begin{minipage}{.32\linewidth}
\centering
\includegraphics[width=1.2\linewidth]{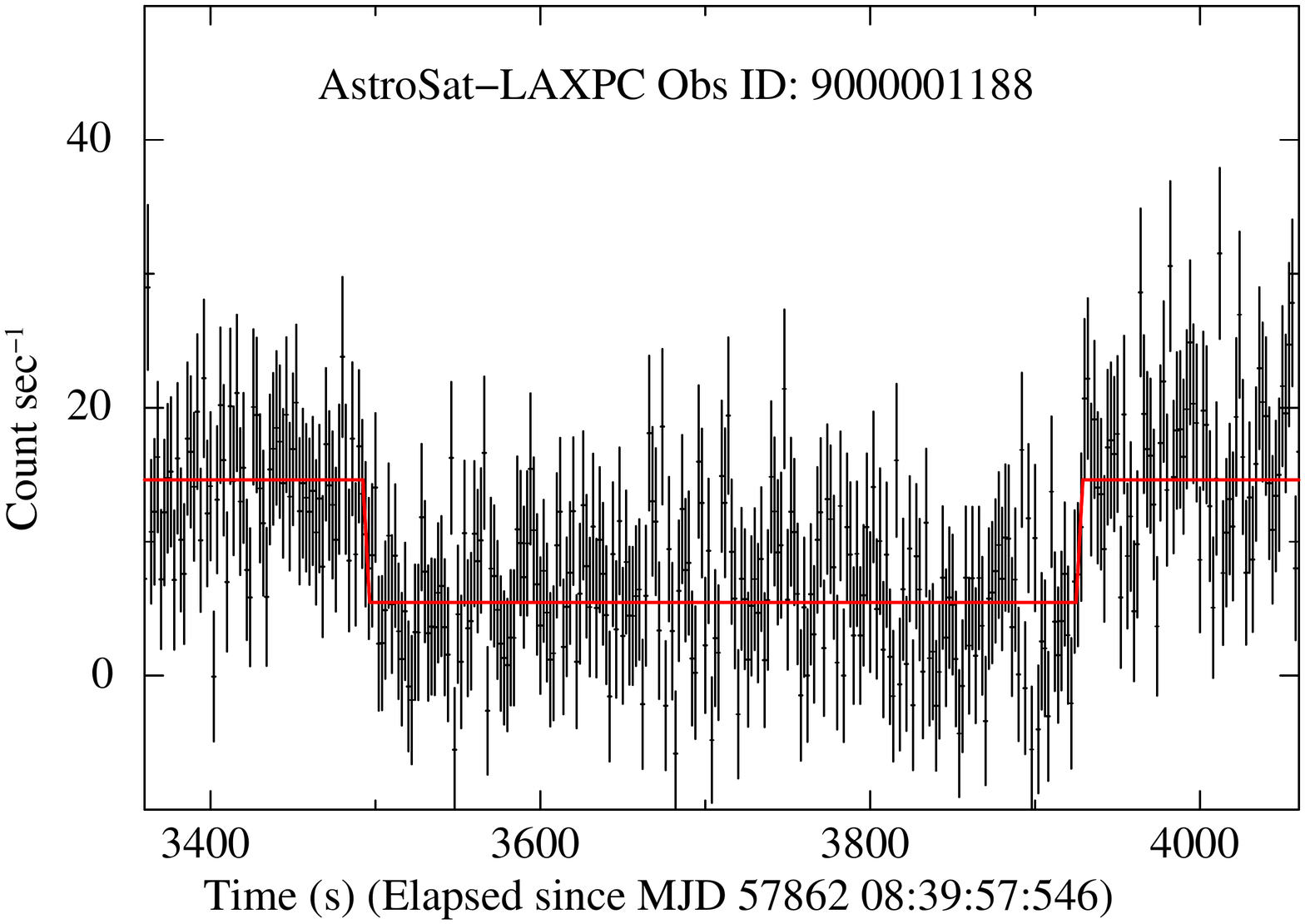}
\end{minipage}
\\[1ex]
\begin{minipage}{.32\linewidth}
\centering
\includegraphics[width=1.2\linewidth]{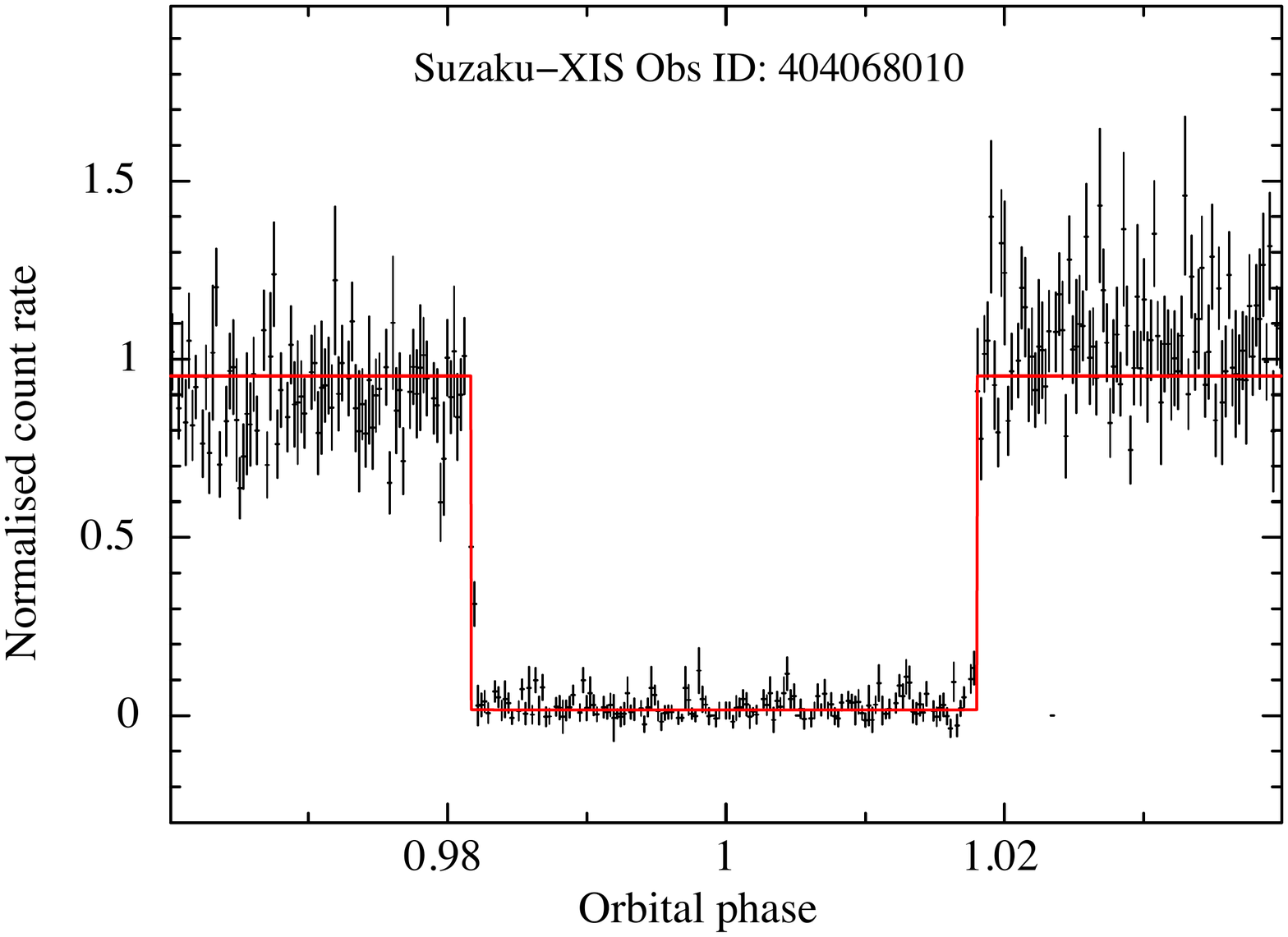}
\end{minipage}
\begin{minipage}{.32\linewidth}
\centering
\includegraphics[width=1.2\linewidth]{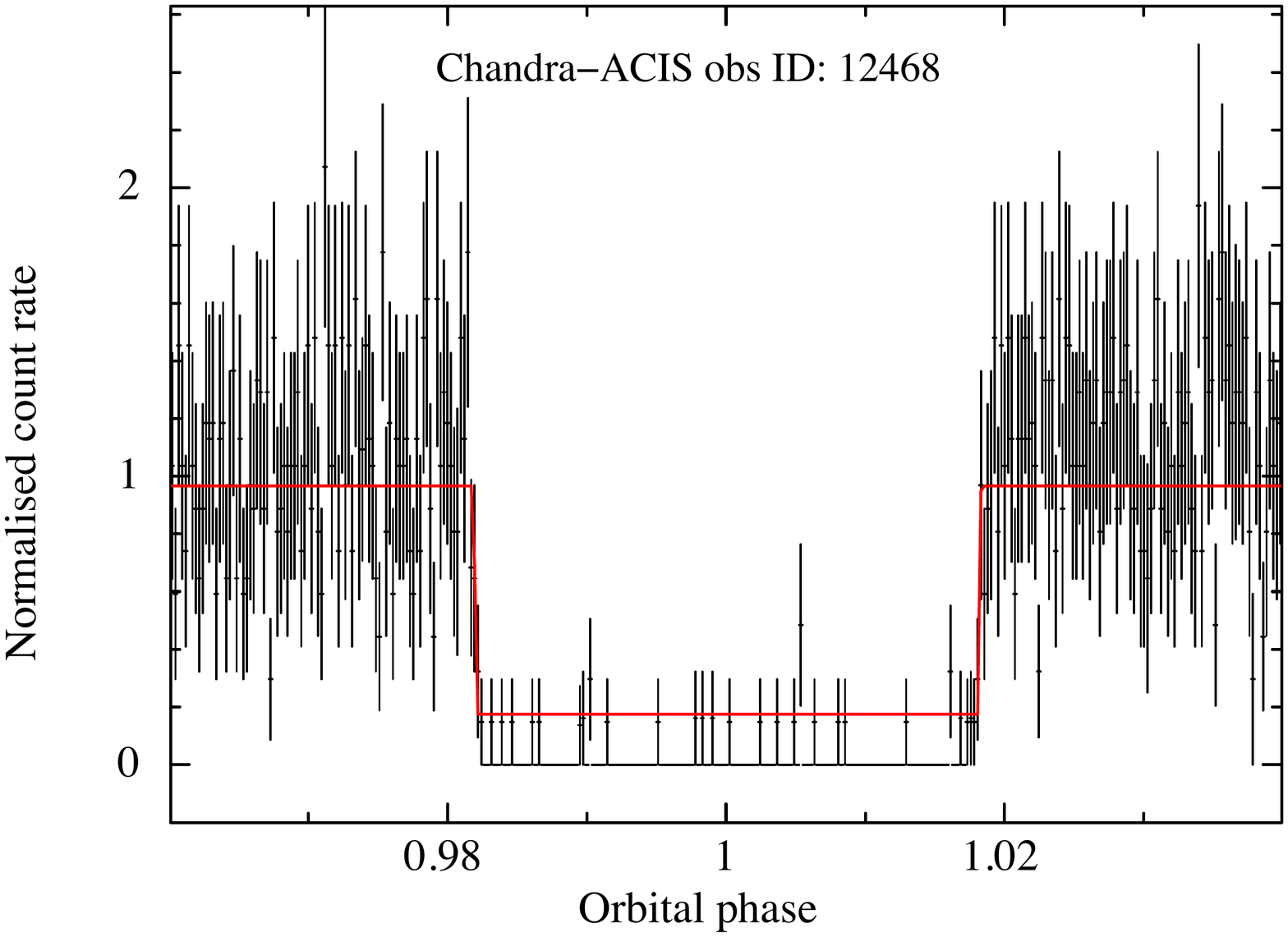}
\end{minipage}
\caption{The eclipse profile of background subtracted light curve of \xte\ obtained from \xmm, \rxte, \astro, \suz\ and \chan. The solid line in each plot represents the best fit five parameter eclipse model.}
\label{fig:eclipse}
\end{figure*}

From data spread over $\sim$ 13 years (2004--2017), we have found 21 complete eclipses (1 with \xmm, 1 with \suz, 13 with \rxte, 2 with \chan\ and 4 with \astro). Figure \ref{fig:eclipse} shows the eclipse phase of background subtracted light curves obtained from \xmm, \suz, \rxte, \chan\ and \astro. For the observations of \xmm-PN, \rxte-PCA and \astro-LAXPC mentioned in Table \ref{table:log}, all the eclipses were analyzed individually. But the light curves from CCD detectors onboard \suz-XIS and \chan-ACIS have been folded to obtained a single eclipse profile. The \suz-XIS and \chan-ACIS light curves were folded with the period mentioned in \citet{Jain11}. The folded orbital profiles from \suz\ and \chan\ data are shown in Figure \ref{fig:eclipse} with respect to respective mid-eclipse time as the epoch. The \suz\ observation duration was $\sim$190 ks, therefore 16 cycles were folded to obtain the orbital profile. This gives a reasonably good resolution for the orbital profile as well as the measurements of eclipse parameters, even though the normal observation mode of the \suz\ data had a time resolution is 8 s. Similarly, for the \chan\ data, about six cycles were folded to obtain the orbital profile. And looking at a time resolution of 1.74 s, it gives a fairly good sensitivity in the measurement of eclipse parameters. 

In order to determine the mid-eclipse times, we fitted an eclipse profile to each eclipse phase. As seen in our earlier paper \citep{Jain11}, for all the observations, the out-of-eclipse count rate did not seem to have any significant variability. It was found that the values of pre--ingress and post--egress count rate were similar and the eclipse ingress and egress duration ($\sim$ 20 s) were also similar within errors. The parameter space for the eclipse model, thus consisted of, (i) The mid-eclipse time, (ii) The eclipse duration, (iii) The ingress transition, (iv) The egress transition and (v) The pre-ingress and the post-egress count rate. In all the eclipse profiles, the eclipse duration was consistent and was observed to last for about 420 s, excluding the ingress and egress transition. We took $\sim$150 s of data before and after the eclipse phase to fit the eclipse profile. 

A sample of the best fit eclipse model from all the five observatories is shown with a solid line in Figure \ref{fig:eclipse}. The mid-eclipse times and the corresponding 1$\sigma$ errors were determined for all the 21 X-ray eclipse profiles. The results are given in Table \ref{table:log}. 
In this table, the orbital cycle is in concurrence with \citet{Jain11}. 

We combined our measurements with the previous measurements \citep{Jain11}. Out of a total 78 eclipse measurements, we have 56 observations during epoch 2 (MJD 52132 - 54410, labelled in Figure \ref{fig:OC}). To determine the secular change in orbital period (other than the glitches) we fitted a constant and a linear model to the eclipse measurements in epoch 2. We obtained an updated constant orbital period of 0.1367109674 (2) d with $1\sigma$ limits of $-1.8 \times 10^{-12}$ d d$^{-1}$ and $0.07 \times 10^{-12}$ d d$^{-1}$ on the period derivative. The best fit model had a reduced $\chi^2$ of 1.9 for 50 degrees of freedom. These limits imply that other than the glitches, the orbital period is either decaying at a timescale $\left(\frac{P_{\rm orb}}{\dot{P}_{\rm orb}}\right)$ of less than $2.1 \times 10^8$ yr or it is increasing at a timescale of $53.5 \times 10^8$ yr. 

\subsection{Case I: Two Orbital Glitches -- Three Epochs}

\begin{figure*}
\centering
\includegraphics[width=0.45\linewidth]{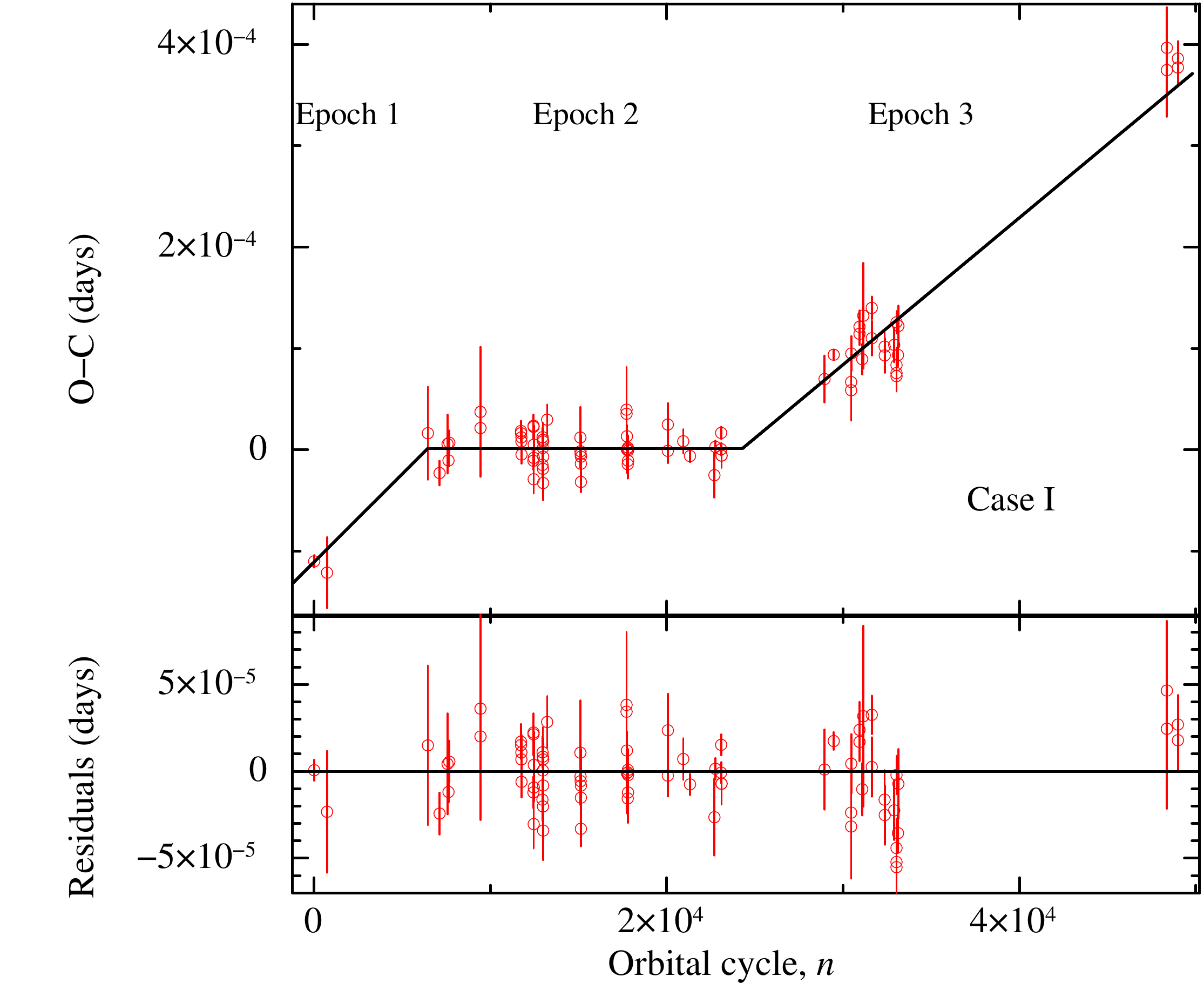}
\includegraphics[width=0.45\linewidth]{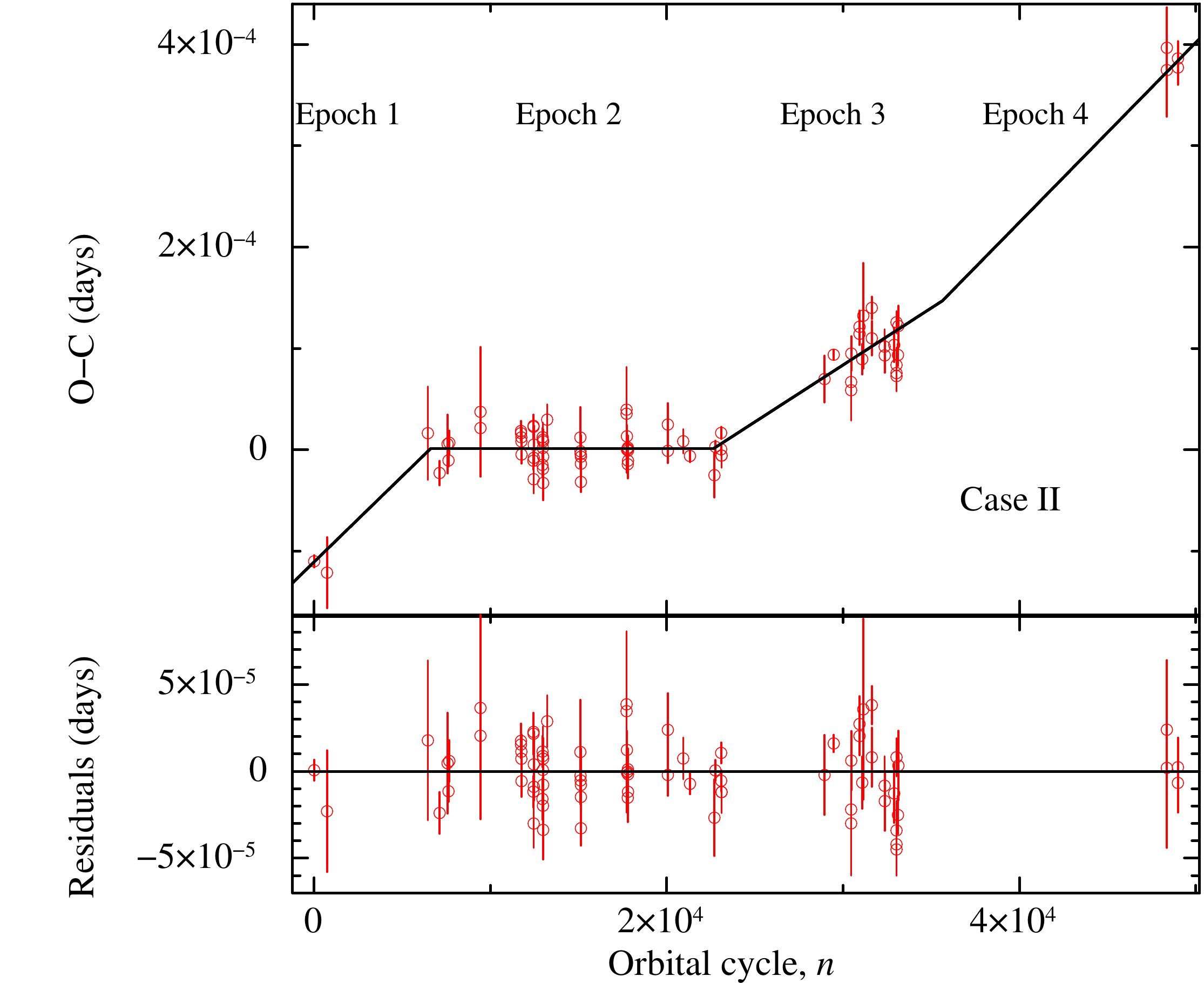}
\caption{\enquote{Observed minus Calculated (O-C)} diagram of \xte\ with two orbital period glitches (Case I) and three orbital period glitches (Case II). The bottom panel in both the diagrams displays the residuals from the best-fit model.}
\label{fig:OC}
\end{figure*}

Figure \ref{fig:OC} shows the \enquote{Observed minus Calculated} (O--C) diagram for all the eclipse measurements of \xte, obtained after subtracting the linear component obtained from epoch 2. The lower panel of the figure shows the residuals from the fitted model. Similar to the observation by \citet{Jain11}, we have found that the O-C diagram cannot be described with a third-order Taylor polynomial. We therefore fitted a piece-wise linear function to the O--C diagram. The best fit model had a reduced $\chi^2$ of 2.5 for 74 degrees of freedom. Due to limited number of data points during epoch 1, we have obtained a lower limit on orbital period change ($\Delta P$) of 1.5 ms between epoch 1 and epoch 2. It is evident from Figure \ref{fig:OC} (left panel), that the second orbital period glitch occurred around orbital cycle 24301 which corresponds to MJD 54573. The results of the fit are given in Table \ref{table:orb_par}.

\subsection{Case II: Three Orbital Glitches; Four Epochs}

From a careful review of Figure \ref{fig:OC} (left panel), it appears that on the O-C diagram, the four measurements from \astro\ observations of 2017 lie significantly away from the best fit solution. We therefore fitted the O--C diagram for \xte\ with piece-wise linear function comprising of four epochs. The best fit functional form along with the residue from the fitted model is shown in Figure \ref{fig:OC} (right panel). The best fit model had a reduced $\chi^2$ of 2.2 for 72 degrees of freedom. As per this model, the second and third glitch occurred around orbital cycle of 22639 and 35639. This corresponds to MJD 54345 and 56123 respectively. 

Following \citep{Jain11}, the O-C values are shown with respect to the second epoch. In order to determine the detection significance of the fourth epoch w.r.t. the second epoch, we fitted a constant to residuals of the fourth epoch. The value of the best-fit constant was divided by the quadrature sum of 1$\sigma$ error associated with the constant fit and the standard deviation of the three glitch model. This gave us a detection significance of 5$\sigma$ for the fourth epoch.

From Table \ref{table:orb_par} it is clear that there is a marginal improvement of about 23 in the value of $\chi^2$ from results of case I to those of case II for two additional parameters. The statistical significance of the three glitch model over the two glitch model can be ascertained from the fact that we have obtained an F-test false alarm probability of $\sim$0.7\% which corresponds to a confidence level of 2.7 $\sigma$. However, given the insufficient number of data points between epoch 3 and epoch 4, it is difficult to determine the exact epoch of the third glitch. As mentioned in Table \ref{table:orb_par}, we report limits of MJD 55726 -- 56402 on the occurrence of the third orbital period glitch.

\begin{figure*}
\centering
\includegraphics[width=0.45\linewidth]{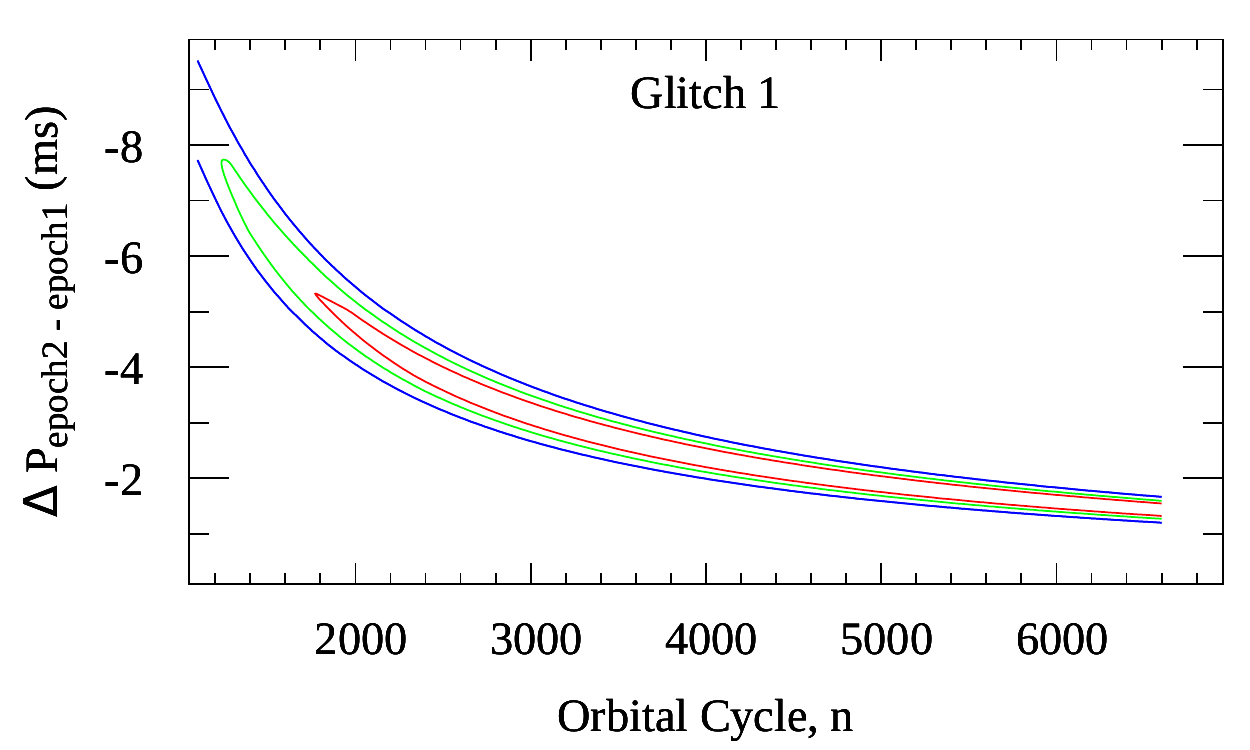}
\includegraphics[width=0.45\linewidth]{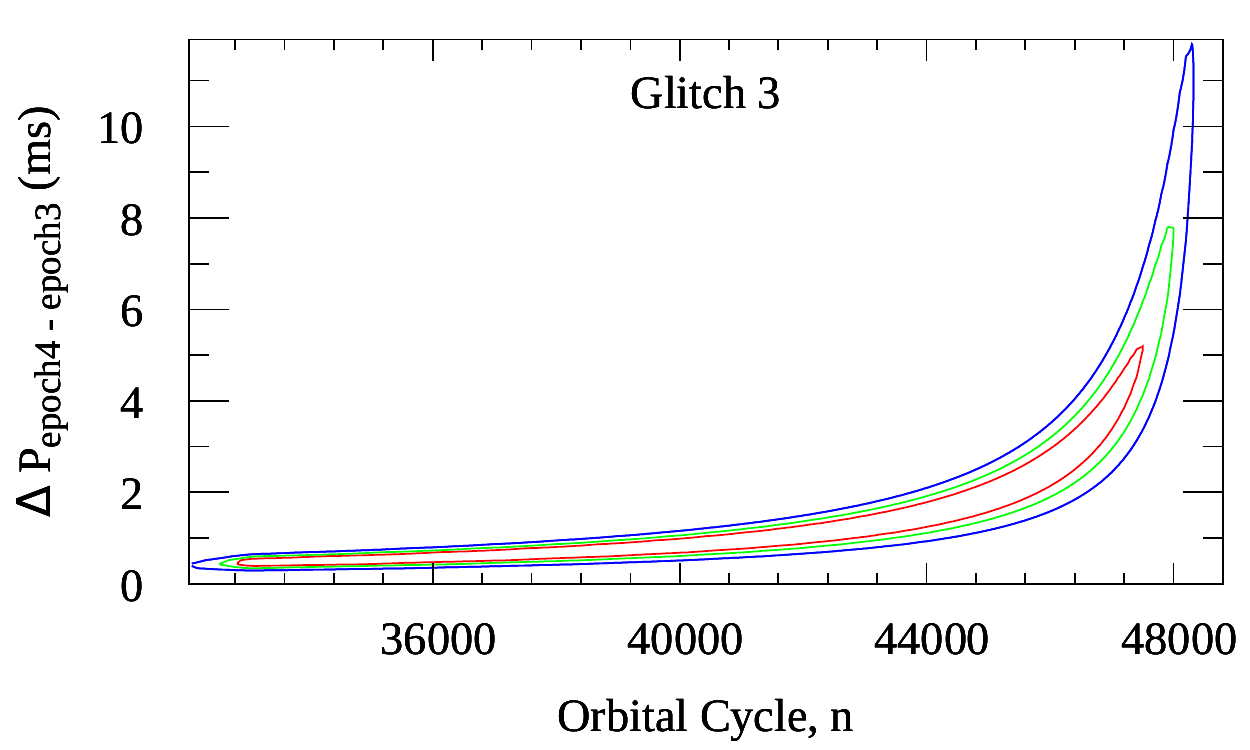}
\caption{Correlation of magnitude of $\Delta P$ with orbital cycle corresponding to the occurrence of first and third glitch. The red, green and blue contours in both the graphs corresponds to $68 \%$, $90 \%$ and $99 \%$ confidence level, respectively.} 
\label{fig:contour}
\end{figure*}

Continuing with the statistical model validation, we have also determined the correlation of change in orbital period $(\Delta P)$ with orbital cycle corresponding to the occurrence of the first and third glitch. This correlation is shown in Figure \ref{fig:contour}. In both the graphs of this figure, the contour plot between orbital cycle and $\Delta P$ is shown for $68 \%$ (red), $90 \%$ (green) and $99 \%$ (blue) confidence level. Clearly, the magnitude of change in the orbital period is correlated with epoch of occurrence of glitch. This figure also establishes that $\Delta P$ reported in this work and in \citet{Jain11} represents the lower limit of $\Delta P$ only. 

Another important inference from the best fit model is the fact that during the first glitch, the orbital period decreased by about 1.48 ms. After about 5 years, the orbital period increased by about 0.97 ms and it increased by about 0.45 ms after a gap of about 6 yr. 

\begin{table}
\centering
\caption{Updated orbital ephemerides of \xte. All the uncertainties quoted in this table are at 90\% confidence level. In this table, $n$ refers to orbital cycle.}
\label{table:orb_par}
\resizebox{0.9\linewidth}{!}{
\begin{tabular}{l l r} 
\hline
Parameter                       &   Case I                      & Case II           \\
\hline
$n_{\rm glitch 1}$              &   \multicolumn{2}{c}{6452*}                     \\
 epoch$_{\rm glitch 1}$ (MJD)   &   \multicolumn{2}{c}{52133*}                     \\
$n_{\rm glitch 2}$              &   24301 (807)                 & 22639 (585)       \\
 epoch$_{\rm glitch 2}$ (MJD)   &   54573 (110)                 & 54345 (80)        \\
$n_{\rm glitch 3}$              &   -                           & 35639$^{+2048}_{-2900}$ \\
epoch$_{\rm glitch 3}$ (MJD)    &   -                           & 55726--56402       \\
$P_{\rm epoch 2} - P_{\rm epoch 1}$ (ms)  &  -1.50 (13)         & -1.48 (13) \\
$P_{\rm epoch 3} - P_{\rm epoch 2}$ (ms)  &  1.24 (9)           & 0.97 (7) \\
$P_{\rm epoch 4} - P_{\rm epoch 3}$ (ms)  &  -                  & 0.45 (15) \\
\hline
$\chi^2_\nu$/dof    &   2.5/74    & 2.2/72 \\
\hline
\multicolumn{3}{L{7cm}}{$*$Although fitting the first glitch is expected to yield a realistic estimation, but owing to only two data points during the first epoch, it will be impractical to mention the error in the estimation of $n_{\rm glitch 1}$ and epoch$_{\rm glitch 1}$ \citep{Jain11}.} \\
\end{tabular}}
\end{table}

\section{DISCUSSION}

This work reports measurement of 21 new mid-eclipse times in \xte\ spread across 13 years. Our results have increased the total number of mid-eclipse time measurements in \xte\ to 78 spanning about 49000 binary orbits. By using data obtained from observations of \xmm, \suz, \rxte, \chan\ and \astro, we have discovered occurrence of a third orbital period glitch in \xte.

The orbital period in \xte\ shows a decrease as well as an increase. As a result, the orbital period glitches have net change of 0.06 ms in about 20 yr timeline. This implies a net period derivative of $0.09 \times 10^{-12}$ d d$^{-1}$. Even if the glitches had same direction, the net period derivative would have been $4.6 \times 10^{-12}$ d d$^{-1}$. In the following sub-sections, we have explored the various possible mechanism for the long-term orbital evolution in \xte\ along with their caveats.

\subsection{Stellar Magnetic Convection}

On the probable cause of orbital period glitches, \citet{Wolff09} had proposed magnetic activity associated with the secondary star as the likely cause in EXO 0748--676. This source was discovered in 1985 and it went into quiescence after about 24 year long X-ray outburst \citep{Parmar85, Degenaar09}. Thus through optical emission, it has been possible to map the magnetic activity in the companion star thereby giving clues on sudden change in the orbital period \citep{Applegate87,Hertz97} in this binary system.

On the other hand, \xte\ is a persistent system where the accretion disc dominates the optical emission. The companion star is expected to become visible only during quiescence. And hence, the magnetic nature of the companion star in \xte\ has not been investigated yet \citep{Ratti10}.

\subsection{Gravitational Wave Radiation}

The rate at which the orbital period shrinks due to loss of energy and angular momentum from a binary system to gravitational waves, is of the order of 10$^{-13}$ d d$^{-1}$ \citep{Pac67, Verbunt93}. Considering gravitational wave radiation as the only cause for orbital period evolution, the orbital period decays ($\dot{P}_{orb}$) according to Equation \ref{eqn1} \citep{Landau71, Ergma99}

\begin{equation} \label{eqn1}
\dot{P}_{orb} = -\frac{192 \pi}{5}\left( \frac{2\pi}{P_{orb}}\right)^{5/3} M_c M_{ns} M^{-1/3} \left( \frac{GM_\odot}{c^3}\right)^{5/3}
\end{equation}

For \xte, mass of neutron star ($M_{ns}$) can be taken as 1.4 $M_{\odot}$. The mass of the companion star ($M_c$) is not known. But assuming it to be a low mass companion with mass in the range 0.01 $M_{\odot}$ to 1 $M_{\odot}$, the total binary mass ($M$) becomes 1.41 $M_{\odot}$ - 2.4 $M_{\odot}$. Plugging in an orbital period ($P_{orb}$) of 0.1367109674 d in Equation \ref{eqn1} gives $\dot{P}_{orb} \sim -(0.07 - 6) \times 10^{-13}$ d d$^{-1}$. From the orbital period glitches, we have estimated a net period derivative of $0.9 \times 10^{-13}$ d d$^{-1}$. Therefore, orbital period glitches are definitely an important clue for understanding the orbital evolution mechanism in \xte\ but looking at the uncertainty in the companion mass, they are certainly not the sole factor responsible for the long term orbital decay. 

\subsection{Hierarchical triple system}

The presence of a third object in an orbit around a binary system is known to alter the evolution of the binary system. Although there are no direct observational clues, nevertheless there are reported works that have established that presence of a massive third object orbiting a binary system is capable of shrinking and expanding its orbital period \citep{Iaria15, Getley17, Jain17}.
In case of \xte, the occurrence of distinct glitches rules out the possibility of triple hierarchical system because the changes in orbital period are expected to evolve continuously as a function of the orbital phase of the third body.

\subsection{Orbital Scattering}

Possible detection of positive and negative orbital period glitch in \xte\ opens up several interesting scenarios and challenges our understanding of the environments of the binary systems. Similar to the solar system, if there are ensemble of smaller bodies and occasional scattering of these bodies by the binary takes place (like comets in the solar system), then that can cause sudden changes in angular momentum and this process can give either a positive or a negative glitch. 

Taking $a$ as the binary separation, the total orbital angular momentum ($J$) is given by Equation \ref{eqn2}.

\begin{equation} \label{eqn2}
J = M_{ns} M_c \left(\frac{G a}{M}\right)^{1/2} = M_{ns} M_c G^{2/3} \left( \frac{P_{orb}}{2 \pi M}  \right)^{1/3}
\end{equation}

Due to impact of an object of mass $m$, the total angular momentum of the system (given in Equation \ref{eqn2}) will change owing to the fact that the total system mass will increase to $M_{ns}+M_c+m$ and the binary period will increase by $\Delta P$. 
This implies that the change in angular momentum due to impact of third object will be (Equation \ref{eqn3})

\begin{equation} \label{eqn3}
\Delta J = \frac{M_{ns} G^{2/3}}{(2\pi)^{1/3}} \left[ (M_c +m) \left( \frac{P_{orb}+\Delta P}{M + m} \right)^{1/3}- M_c \left( \frac{P_{orb}}{M}\right)^{1/3} \right]
\end{equation}

Assuming that the object of mass $m$ is falling freely onto the binary and considering extreme case that this object is captured by the companion star, the change in the total angular momentum of the binary $(\Delta J)$ will be given by Equation \ref{eqn4}.

\begin{equation} \label{eqn4}
\Delta J = m \sqrt{2GMa}
\end{equation}

Taking extreme range of 0.01 -- 1 $M_{\odot}$ for the companion mass and simultaneously solving Equations \ref{eqn3} and \ref{eqn4}, for $P_{orb} \sim 11833$ s, $\Delta P \sim 1.5$ ms, and $M_{ns}$ = 1.4 $M_{\odot}$, the mass of the third object interacting with the binary system is expected to be $\approx 10^{23-24}$ g.

Considering our solar system, although the cometary mass depends on its size and the mean density of its nucleus, nevertheless, the most massive comets are known to have mass in the range $10^{17}-10^{19}$ g \citep{Hughes85, Hughes90}. But these comets are not known to induce a significant perturbation in the orbit of solar system satellites. So any hypothesis of labelling the third object in the \xte\ system as an exo-comet needs a strong argument.

It is known that a significant fraction of white dwarfs harbour a large number of extra-solar minor bodies having mass in the range of 10$^{19}$ to 10$^{26}$ g \citep{Farihi10, Veras14, Veras16, Strom20}. These minor objects are often considered equivalent to highly eccentric asteroids which get tidally disrupted around the compact object \citep{Campana11}. In fact, long term monitoring of several neutron star systems has indicated presence of an asteroid belt around pulsars \citep{Shannon13, Brook14}. There are also cases where debris in the supernova fallback accretion disk around a neutron star gets periodically perturbed by a large orbiting object (likely to be a planet) \citep{Cordes08}. 

In case of \xte, looking at the probable mass of the third body, it could be possible that extra-solar planetesimals are present around the binary system. But it is too early to comment on the viability of this conjecture. The results presented in this work are encouraging to plan future X-ray observations during the current active phase to refine the occurrence of orbital period glitch, to explore the possibility of another glitch, to investigate detailed/ alternative interpretation of the cause for occurrence of orbital glitch and thereby monitor the orbital evolution in \xte. It will also be useful to carry out optical observations of the companion star during a future quiescent phase (if any).

\section*{Acknowledgments}
This work has made use of data from the \astro\ mission of the Indian Space Research Organisation, archived at the Indian Space Science Data Centre. We thank the LAXPC Payload Operation Center (POC) at TIFR, Mumbai for providing necessary software tools. This work has also made use of data and software provided by the High Energy Astrophysics Science Archive Research Center (HEASARC), which is a service of the Astrophysics Science Division at NASA/GSFC. We also thank the anonymous referee for insightful comments and suggestions.


\section*{Data Availability}

Data used in this work can be accessed through the Indian Space Science Data Center at 
\url{https://astrobrowse.issdc.gov.in/astro\_archive/archive/Home.jsp}, and HEASARC archive at \url{https://heasarc.gsfc.nasa.gov/cgi-bin/W3Browse/w3browse.pl}.



\bibliographystyle{mnras}
\bibliography{refs} 


\bsp	
\label{lastpage}
\end{document}